\documentclass[aoas]{imsart}

\RequirePackage{amsthm,amsmath,amsfonts,amssymb}
\RequirePackage[authoryear]{natbib}
\RequirePackage[colorlinks,citecolor=blue,urlcolor=blue]{hyperref}
\RequirePackage{graphicx}
\usepackage{multirow}
\usepackage{enumerate}
\usepackage{enumitem}
\usepackage{xcolor}
\usepackage{color}
\usepackage{url}
\usepackage{colortbl}
\usepackage{bm}
\usepackage{graphics}
\usepackage{epstopdf}
\usepackage[T1]{fontenc}
\usepackage[font=small,labelfont=bf,tableposition=top]{caption}
\usepackage{threeparttable}
\usepackage{booktabs}
\usepackage{array}
\usepackage{epsfig}
\usepackage{mathrsfs}
\usepackage{centernot}
\usepackage{subcaption}
\usepackage{algorithm}
\usepackage{algorithmic}
\usepackage[export]{adjustbox}

\startlocaldefs
\theoremstyle{plain}
\newtheorem{theorem}{Theorem}

\theoremstyle{definition}  
\newtheorem{definition}{Definition}

\newcommand{\ctheta}{\mu_{XY|Z}}

\newcommand{\cmu}{\mu_{X|Z}}
\newcommand{\cnu}{\mu_{Y|Z}}

\newcommand{\CI}{{\perp\mspace{-10mu}\perp}}

\newcommand{\XYCI}{X\CI Y | Z}

\newcommand{\bally}{\bar{B}_{\mathcal{Y}}(y_1,y_2)}

\newcommand{\sisstat}{\hat{\tau}}

\newcommand{\ijklst}{ijklst}

\endlocaldefs

\begin{document}

\begin{frontmatter}
\title{Identification of Genetic Factors Associated with Corpus Callosum Morphology: Conditional Strong Independence Screening for Non-Euclidean Responses}
\runtitle{Conditional Strong Independence Screening for Non-Euclidean Responses}

\begin{aug}
\author[A]{\fnms{Zhe} \snm{Gao}\ead[label=e1,mark]{gaozh8@mail.ustc.edu.cn}},
\author[B]{\fnms{Jin} \snm{Zhu}\ead[label=e2,mark]{j.zhu.7@bham.ac.uk}},
\author[C]{\fnms{Yue} \snm{Hu}\ead[label=e3]{yue.hu@yale.edu}},
\author[D]{\fnms{Wenliang} \snm{Pan}\ead[label=e4,mark]{panwliang@amss.ac.cn}}
\and
\author[A]{\fnms{Xueqin} \snm{Wang}\ead[label=e5]{wangxq20@ustc.edu.cn}}
\address[A]{School of Management, University of Science and Technology of China, \printead{e1,e5}}
\address[B]{School of Mathematics, University of Birmingham, \printead{e2}}
\address[C]{Yale University School of Public Health \printead{e3}}
\address[D]{State Key Laboratory of Mathematical Sciences, Academy of Mathematics and Systems Science, Chinese Academy of Sciences \printead{e4}}
\end{aug}

\begin{abstract}
   
    The corpus callosum, the largest white matter structure in the brain, plays a critical role in interhemispheric communication. Variations in its morphology are associated with various neurological and psychological conditions, making it a key focus in neurogenetics. Age is known to influence the structure and morphology of the corpus callosum significantly, complicating the identification of specific genetic factors that contribute to its shape and size. We propose a conditional strong independence screening method to address these challenges for ultrahigh-dimensional predictors and non-Euclidean responses, incorporating prior knowledge such as age through a novel concept of conditional metric dependence, which quantifies nonlinear conditional dependencies among random objects in metric spaces without relying on predefined models. We apply this framework to identify genetic factors associated with the morphology of the corpus callosum. Simulation results demonstrate the efficacy of this method across various non-Euclidean data types, highlighting its potential to drive genetic discovery in neuroscience.
    
\end{abstract}

\begin{keyword}
\kwd{Conditional metric dependence}
\kwd{Conditional strong independence screening}
\kwd{Non-Euclidean response}
\kwd{Ultrahigh-dimensional data analysis}
\end{keyword}

\end{frontmatter}

\section{Introduction}

Neuroscience has seen transformative advancements, moving from basic anatomical studies to highly sophisticated computational analyses. The advent of neuroimaging technologies, such as magnetic resonance imaging (MRI), diffusion tensor imaging (DTI), and positron emission tomography (PET), has revolutionized the way we visualize and understand the brain's complex structures. These innovations have deepened our insights into neural anatomy and function while introducing non-Euclidean data that require novel analytical methods. One prominent example is the corpus callosum, the brain’s largest white matter structure. Comprising approximately 200 million myelinated nerve fibers, it is the primary connection between the left and right hemispheres \citep{tanaka2015developmental}. These fibers form homotopic and heterotopic projections, enabling the integration and transfer of sensory, motor, and cognitive information between hemispheres \citep{kennedy2016micro}. Changes in the size and shape of the corpus callosum can significantly impact its function and have been associated with several neurological conditions, including Alzheimer's disease \citep{bachman2014corpus}, schizophrenia, and bipolar disorder \citep{vermeulen2023morphological}.

The goal of genome-wide association studies (GWAS) that look into brain structure is to find out how genetic factors affect the structure and function of the brain. Researchers often collect high-dimensional genetic data and neurological metrics to do this. Along with genetic information, covariates like age and disease status are often included because they significantly affect brain structure \citep{cornea2016regression, biswal2010toward}. However, the specific genetic factors that shape the morphology of the corpus callosum remain largely unexplored. Given the critical role of covariates like age in influencing corpus callosum structure, assessing the conditional contributions of genetic factors while adjusting for these covariates to identify key genetic determinants is essential. Identifying relevant factors from high-dimensional data while accounting for essential covariates remains crucial in analyzing non-Euclidean data.

To address the challenge of high-dimensional data analysis, \citet{barut2016csis} introduced conditional sure independence screening (CSIS) within the generalized linear model framework, expanding upon the original sure independence screening (SIS) method \citep{fan2008sure}. CSIS evaluates the conditional maximum likelihood estimator to measure marginal utility and selects variables with high marginal utility. Since its inception, CSIS has been adapted to various contexts. For instance, \citet{hu2017conditional} developed an empirical-likelihood-based CSIS for less restrictive distributional assumptions, while \citet{hong2018conditional} introduced a version for censored responses. Furthermore, \citet{chen2019simple} suggested a model-free version, and \citet{wen2018sure} made a conditional distance correlation-based CSIS for responses with multiple variables. CSIS has also been extended to survival data \citep{lu2019conditional}. Other changes include an adaptive CSIS created by \citet{lin2015adaptive} to cut down on false positives and negatives, as well as robust CSIS methods for weaker model assumptions created by \citet{hong2016data} and \citet{zhang2018robust}. Furthermore, \citet{liu2018quantile} suggested a conditional quantile independence screening method to find features that affect the conditional quantiles of the response without defining a model structure, and \citet{zheng2020building} introduced a sequential conditioning method that updates the conditioning set over and over again.

Despite these numerous extensions, none of the existing CSIS methods are designed explicitly for non-Euclidean responses. This gap underscores the need for new approaches to handle the complex data structures encountered in modern neurogenetic studies.

This article aims to explore genetic factors influencing the shape of the corpus callosum by introducing a novel conditional screening method that addresses these gaps. Methodologically, we propose a model-free and distribution-free procedure called Conditional Metric Dependence-based Conditional Strong Independence Screening (COME-CSIS). Theoretically, COME-CSIS has been proven effective across a wide range of non-Euclidean data and exhibits strong screening properties in ultra-high-dimensional settings. In addition, COME-CSIS offers several significant advantages. It is robust, allowing for less restrictive moment conditions on both explanatory variables and the response. Its corresponding marginal utility enables conditional dependence analysis in non-Euclidean spaces, which has remained largely unexplored.

By applying COME-CSIS to both synthetic and real datasets, we demonstrate its ability to overcome the limitations of the strong negative type \citep{lyons2013distance}, establishing it as a valuable tool for ultrahigh-dimensional non-Euclidean data analysis. In the real-data application, COME-CSIS was employed to identify genes associated with the shape of the corpus callosum. Compared to existing screening methods, our approach avoided selecting genes related primarily to age or Alzheimer's disease, which are well-known factors influencing corpus callosum morphology. These results provide important preliminary insights into the gene regulation mechanisms affecting the corpus callosum and open the door for further investigation.

The following contents are organized as follows:
In Section~\ref{sec:preliminary}, we introduce notation and formulate our problem.
Then, we propose a novel marginal utility for CSIS and a related CSIS procedure
and study their theoretical properties in Section~\ref{sec:COME-CSIS}.
In Section~\ref{sec:Emp}, we conduct simulation studies to evaluate the performance of our method.
In Section~\ref{sec:App}, we analyze a real-world dataset to demonstrate the helpfulness of our method.
Finally, we conclude this article with a few remarks in Section~\ref{sec:dis}.
All of the technical details are deferred to the supplementary materials.

\section{Preliminaries}\label{sec:preliminary}
The section will introduce notations and formulate our problem in subsection~\ref{subsec:notation} and \ref{subsec:formulation}, respectively.

\subsection{Notations}\label{subsec:notation}

Let $(X, Y, Z)$ be a random variable defined on a probability space, where 
$Y \in \mathcal{Y}$ is a metric-valued variable, and $X, Z$ are Euclidean-valued variables with respective dimensionalities $d$ and $d_z$. We decompose $X$ into $p$ variables, denoted as $X = (X_1, \ldots, X_p)$, where each component $X_i$ has dimensionality $d_i$, satisfying $\sum_{i=1}^p d_i = d$. Assume that the regular conditional probabilities of $(X_1, Y), \ldots, (X_p, Y)$, and $(X_1, \ldots, X_p, Y)$ given $Z$ exist.

To facilitate the analysis of the variable $Y$ in the non-Euclidean space $\mathcal{Y}$, we introduce the following notation and assumptions. Suppose $(\mathcal{Y}, \rho)$ is a Polish space, endowed with a metric $\rho$. Let  $\bar{B}_{\mathcal{Y}}(u, v)$ represents a closed ball centered at $u$ with radius $r = \rho(u, v)$. An analogous definition applies to the closed balls in the Euclidean space $\mathbb{R}^d$ where $X$ is located. The product metric space is defined as $(\mathbb{R}^d \times \mathcal{Y}, \rho_{\mathbb{R}^d \times \mathcal{Y}})$. Denote the regular conditional probabilities of $(X, Y), X, Y$ given $Z$ as $\ctheta, \cmu, \cnu$. 

We further restrict the space $(\mathcal{Y}, \rho)$ by introducing the following definition:
\begin{definition}[\cite{federer2014geometric}]
    A metric $\rho$ is called directionally $(\epsilon, \eta, L)$-limited at the subset $A$ of $\mathcal{Y}$, if $\epsilon>0,0<\eta \leq 1 / 3, L$ is a positive integer, and the following condition holds: if for each $a \in A, D \subseteq A \cap \bar{B}_{\mathcal{Y}}(a, \epsilon)$ such that $\rho(x, c) \geq \eta \rho(a, c)$ whenever $b, c \in D, b \neq c, x \in \mathcal{Y}$ with
    $$
    \rho(a, x)=\rho(a, c), \rho(x, b)=\rho(a, b)-\rho(a, x)
    $$
then the cardinality of $D$ is no larger than $L$.
\end{definition}
\noindent Directionally $(\epsilon, \eta, L)$-limited stipulates that the directionality at each local point must be finite. More details can be found in \citep{federer2014geometric,wang2021nonparametric}.

Let $\{(X_{1i}, \ldots, X_{pi}, Y_{i}, Z_{i})\}_{i=1}^{n}$ be an \emph{i.i.d.} sample of $(X_1, \ldots, X_p, Y, Z)$. Suppose $I(\cdot)$ is an indicator function, we define $\delta^{X_r}_{ij,k}=I(X_{rk}\in \bar{B}_{\mathbb{R}^{d_r}}(X_{ri}, X_{rj}))$, 
which indicates whether $X_{rk}$ is located in $\bar{B}_{\mathbb{R}^{d_r}}(X_{ri}, X_{rj})$, 
and $\delta^{X_r}_{ij,kl}=\delta^{X_r}_{ij,k}\delta^{X_r}_{ij,l}$, which is an indicator for whether both of $X_{rk}$ and $X_{rl}$ fall into $\bar{B}_{\mathbb{R}^{d_r}}(X_{ri}, X_{rj})$. 
Also, let $\eta^{X_r}_{ij,klst}=(\delta^{X_r}_{ij,kl}+\delta^{X_r}_{ij,st}-\delta^{X_r}_{ij,ks}-\delta^{X_r}_{ij,lt})/2$.
Analogously, define the notations $\delta^Y_{ij,k}$, $\delta^Y_{ij,kl}$ and $\eta^Y_{ij,klst}$ for $Y$.
The product of $\eta^{X_r}_{ij,klst}$ and $\eta^Y_{ij,klst}$ is defined as $\zeta_{\ijklst, r} =\eta_{\ijklst}^{X_r}\eta_{\ijklst}^{Y}$.

\subsection{Problem formulation}\label{subsec:formulation}
The primary objective of this article is to establish a conditional sure independence screening method for identifying relevant factors from high-dimensional data, particularly when considering response variables in the analysis of non-Euclidean data. Consider the random objects $(X_1, \ldots, X_p, Y, Z)$ defined previously, where $p$ is large, indicating that only a small fraction of the $X$ variables are relevant to $Y$ given $Z$. We characterize this relevance through conditional independence: if $X_r \perp Y \mid Z$, then the variable $X_r$ is deemed unnecessary. Conversely, if $X_r \not\perp Y \mid Z$, then $X_r$ is considered necessary. Here, $\perp$ and $\not\perp$ denote conditional independence and dependence, respectively. 

We define the inactive set and active set as follows:
\begin{align*}
  \mathcal{I} &= \{r \mid r \in \{1, \ldots, p\} \text{ and } X_r \perp Y \mid Z\}, \\
  \mathcal{A} &= \{r \mid r \in \{1, \ldots, p\} \text{ and } X_r \not\perp Y \mid Z\}.
\end{align*}
Our goal is to accurately identify the set $\mathcal{A}$. Returning to the real-world problem discussed in the previous section, consider $Y$ as the shape of the corpus callosum, and each $X_r$ (for $i = 1, \ldots, p$) as a genetic factor. $Z$ includes confounding variables that are presumed to influence the relationship between the shape of the corpus callosum and genetic factors, such as age, gender, and disease status. Our objective is to detect the genetic factors that are truly associated with the shape of the corpus callosum while accounting for these confounding variables, effectively recovering the set $\mathcal{A}$.

\section{Methods}\label{sec:COME-CSIS}
In this section, we systematically introduce the measure of conditional independence for general spaces and the screening process developed based on this measure. Additionally, we delineate the theoretical properties that substantiate the efficacy of the proposed method. 

\subsection{Conditional metric dependence (COME)}
We will present two types of conditional metric dependence (COME): conditional ball covariance and global conditional ball correlation in this part. 
Our key idea for the COME measures is to evaluate the difference between
$\ctheta$ and the product measure of $\cmu$ and $\cnu$ on the close ball set.
We first present the definitions of conditional ball covariance.

\begin{definition}\label{def:CBCov}
The conditional ball covariance is defined as
\begin{align*}
  \mathcal{V}^2(X,Y|Z)=&\int \big[\ctheta\big(\bar{B}_{\mathbb{R}^d}(x_1, x_2) \times \bally \big) - \cmu\big(\bar{B}_{\mathbb{R}^d}(x_1, x_2) \big) \cnu\big(\bally \big)\big]^2 \\
  &\quad \times \ctheta(dx_1,dy_1)\ctheta(dx_2,dy_2).
\end{align*}
\end{definition}
\noindent Notably, $\mathcal{V}^2(X,Y|Z)$ is a non-negative random variables of $Z$.
It evaluates the conditional dependence between $X$ and $Y$ for all points in the support set of $Z$, and thus, it can be considered as a ``local'' conditional dependence measure.
Similar to the traditional correlation coefficient, we can also define the conditional ball correlation
\begin{definition}\label{def:CBCor}
The conditional ball correlation is defined as
\begin{align*}
  \mathcal{R}^2(X,Y|Z)= \frac{\mathcal{V}^2(X,Y|Z)}{\sqrt{\mathcal{V}^2(X, X|Z)\mathcal{V}^2(Y, Y|Z)}}.
\end{align*}
if $\mathcal{V}^2(X, X|Z)\mathcal{V}^2(Y, Y|Z)> 0$, and 0 otherwise.
\end{definition}

Next, we present a ``global'' measure for conditional dependence by take expectation on the condition variable $Z$.
\begin{definition}\label{def:CBCG}
  The global conditional ball correlation is
  \begin{align*}
    \tau =E[ \mathcal{R}^2(X,Y|Z) ].
  \end{align*}
\end{definition}
\noindent The metrics defined in Definitions \ref{def:CBCov} and \ref{def:CBCG} are pivotal to our analysis and are utilized to quantify the correlation between random variables $X$ and $Y$, conditioned on $Z$. The conditional ball covariance specifically measures local conditional correlations given $Z$. However,  $\mathcal{R}^2(X,Y|Z)$ is a function of $Z$, it is not directly suitable for ranking purposes and may not  be suitable as a marginal utility function in certain scenarios. Back to the previous application problem, when examining the relationship between the shape of the corpus callosum and specific genes, conditioned on age, the global conditional ball correlation is the appropriate metric. Conversely, when the focus is strictly on the population aged 60, conditional ball covariance should be applied directly.

We turn to show that the two COME measures are proper conditional dependence measures under mild conditions. Our first condition is related to probability measures. 
Let $\mathbf{M}_1$ be a collection of discrete Borel probability measures on $\mathbb{R}^d \times \mathcal{Y}$, 
and $\mathbf{M}_2$ be a collection of Borel probability measures on $\mathbb{R}^d \times \mathcal{Y}$ such that
$\forall (x_1,y_1)\in \mathbb{R}^d \times \mathcal{Y}$ and 
$(x_2,y_2) \sim \ctheta,  \big( \rho_{\mathbb{R}^d}(x_1, x_2), \rho_{\mathcal{Y}}(y_1, y_2) \big)$ 
has a continuous density function. 
The first condition is:
\begin{enumerate}[label=(C\arabic*)]
  \item Both $\mu_{X|Z}$ and $\mu_{Y|Z}$ are convex combinations of elements in $\mathbf{M}_1$ and $\mathbf{M}_2$. \label{pm-condition}
\end{enumerate}

\noindent Our second condition relies on the restriction of $(\mathcal{Y}, \rho)$:
\begin{enumerate}[label=(C\arabic*), start=2]
  \item $\rho$ is directionally-$(\epsilon, \eta, L)$ limited at $\mathcal{Y}$.
  \label{space-condition1} 
\end{enumerate}

The essence of the conditional independence measure is based on comparing differences across a set of closed balls. This approach allows our method to be effectively applied in complex spaces, exhibiting desirable performance. The foundational attribute of these desirable properties lies in the finite covering theorem for closed balls. This theorem is akin to the Vitali covering theorem in Euclidean spaces, which is crucial for establishing the correspondence theorem. Condition \ref{space-condition1} is essential to ensure that the covering theorem remains valid in the complex space $\mathcal{Y}$. Specifically, \ref{space-condition1} mandates that directionality at each local point is finite. We further illustrate the applicability of the modified \ref{space-condition1} with examples such as finite-dimensional Banach spaces equipped with a norm, finite-dimensional Riemannian manifolds with bounded sectional curvature, and metric spaces corresponding to binary phylogenetic trees with a fixed number of leaves.
Given that $X$ resides in a Euclidean space, condition \ref{space-condition1} is naturally satisfied. Condition \ref{pm-condition} ensures that, within the framework of a product space, the directionality constraints of each space are sufficient to uphold the covering theorem. Concurrently, sets $\mathbf{M}_1$ and $\mathbf{M}_2$ cover the majority of the measure, which is not an overly stringent assumption.
As far as we know, these two conditions cannot be relaxed, otherwise counterexamples can be given. For a more detailed discussion, please refer to \citep{wang2021nonparametric}.

The theorem below guarantees the COME measures are proper measurements for conditional dependence. 
\begin{theorem}\label{theorem:equivalence}
 If \ref{pm-condition} and \ref{space-condition1} hold, the following properties hold: 
 \begin{enumerate}
  \item[(i)] $\mathcal{V}^2(X,Y|Z)\geq0$ almost surely, and the equality holds if and only if $\XYCI$.
  \item[(ii)] $0 \leq \tau \leq 1$, and $\tau = 0$ if and only if $\XYCI$. 
 \end{enumerate}
\end{theorem}

Theorem 1 shows that, under mild conditions, $\mathcal{V}^2(X,Y|Z) \overset{a.s.}{=} 0$ and $\tau = 0$ 
are equivalent to conditional independence. 
Many well-known conditional independence measures satisfy similar conditional-independence equivalence, 
like the kernel conditional dependence measures \citep{fukumizu2008kernel} and the conditional distance correlation \citep{wang2015conditional}.
Many metric spaces encountered in non-Euclidean data analysis are directionally limited, such as Riemannian manifolds and phylogenetic tree spaces \citep{wang2021nonparametric}.

Theorem 1 offers a theoretical guarantee for our analytical procedures. Specifically, when employing the COME technology, this theorem ensures that irrelevant variables are not detected. In the absence of Theorem 1, the application of COME to analyze the shape of the corpus callosum could inadvertently lead to the selection of genes that have no relevance to the corpus callosum (CC) shape. Such an occurrence would undermine the credibility of our results and pose challenges for subsequent research endeavors.

For practical usage, we propose the COME estimators. 
We first show that the conditional ball covariance is the conditional expectation of two random variables:
\begin{theorem}\label{theorem:separability}
  A variable-separated variant of the conditional ball covariance is given as follows, 
 \begin{align*}
   \mathcal{W}^{X,Y}(z)=E\{\eta^X_{12,3456}\eta^Y_{12,3456}|Z_1, \ldots, Z_6= z\}.
 \end{align*}
\end{theorem}
\noindent Theorem~\ref{theorem:separability} motivates us to estimate $\tau$ with V-processes and density estimation of $Z$.
Denote $H$ and $K(\cdot)$ as bandwidth matrix and kernel function, respectively.
Let $\hat{f}_k(z) = (n|H|)^{-1}K(H^{-1}(z-Z_k))$, the kernel density estimation for $Z = z$ is presented as
$\hat{f}(z)=\sum_{k=1}^n \hat{f}_k(z)$.
Then $\mathcal{W}^{X,Y}(z)$ can be estimated by the empirical V-process:
\begin{align*}
  \mathcal{W}_n^{XY}(z) = \sum_{i, j, k, l, s, t=1}^{n} \frac{\hat{f}_i(z)\hat{f}_j(z)\hat{f}_k(z)\hat{f}_l(z)\hat{f}_s(z)\hat{f}_t(z)}{\hat{f}^6(z)}\eta^X_{\ijklst}\eta^Y_{\ijklst}.
\end{align*}
And $\mathcal{R}^2(X,Y|Z=z)$ can be estimated by: 
\begin{align*}
\mathcal{R}^2_n(X,Y|Z=z)=\mathcal{W}_n^{XY}(z) / \sqrt{ \mathcal{W}_n^{XX}(z) \mathcal{W}_n^{YY}(z)},
\end{align*}
if $\mathcal{W}_n^{XX}(z) \mathcal{W}_n^{YY}(z) > 0$, and 0 otherwise. 
From the definition of $\tau$, 
it can be estimated by
\begin{equation}\label{bcd-statistic}
  \sisstat=\frac{1}{n}\sum_{u=1}^{n} \mathcal{R}^2_n(X,Y|Z=z_u).
\end{equation}
The computation burden would be extremely heavy if computing $\sisstat$ according to its exact definition \eqref{bcd-statistic}.
In supplementary material, we will develop a method to significantly reduce the time complexity of computing $\sisstat$.

\subsection{COME based conditional strong independence screening}
The subsection establishes a novel CSIS procedure to tackle the problem formulated in subsection~\ref{subsec:formulation}. 
According to Theorem~\ref{theorem:equivalence}, under mild conditions, the global ball condition correlation equal to zero is equivalent to the conditional independence 
thereby, we have:
\begin{align*}
  \mathcal{A} = \{r \mid  \tau_r > 0 \} \textup{ and } \mathcal{I} = \{r \mid  \tau_r = 0 \},
\end{align*}
where $\tau_r$ is the global conditional ball correlation between $X_r$ and $Y$ given $Z$.
This fact motivates us to calculate the global conditional ball correlation for pair $(X_r, Y, Z)$ as a marginal utility to assess the influence of $X_r$ on the response $Y$ given $Z$, 
and then filter out the variables with zero global conditional metric dependence.
Based on this motivation, we propose a two-step procedure, COME-based conditional strong independence screening (COME-CSIS):
 
\begin{enumerate}
  \item[(i)] Given a dataset $\{(X_{1i}, \ldots, X_{pi}, Y_{i}, Z_{i})\}_{i=1}^{n}$, we compute $\hat{\tau}_{1}, \ldots, \hat{\tau}_{p}$ according to~\eqref{bcd-statistic};
  \item[(ii)]  Using $\hat{\tau}_{i}$ as a marginal utility, we select the $X_r$ that fall into
  \begin{align*}
    \hat{\mathcal{M}}_{d_n} = \{r \mid \hat{\tau}_{r} > \tau_n, r = 1, \ldots, p \}.
  \end{align*} 
  where $\tau_n$ is a pre-specified constant.
\end{enumerate}
In practical applications, we also recommend using the hard threshold method to truncate $\hat{\tau}_{1}, \ldots, \hat{\tau}_{p}$, that is, select the first $d_n$ variables, where $d_n$ is related to $n$ \citep{fan2008sure}.
 \begin{align*}
    \hat{\mathcal{M}}_{d_n} = \{r \mid \hat{\tau}_{r}\textup{ is the first } d_n \textup{ largest of }\{\hat{\tau}_{1}, \ldots, \hat{\tau}_{p}\} \}.
  \end{align*}

\begin{figure}[htbp]
  \begin{center}
    \includegraphics[width = 0.95\textwidth]{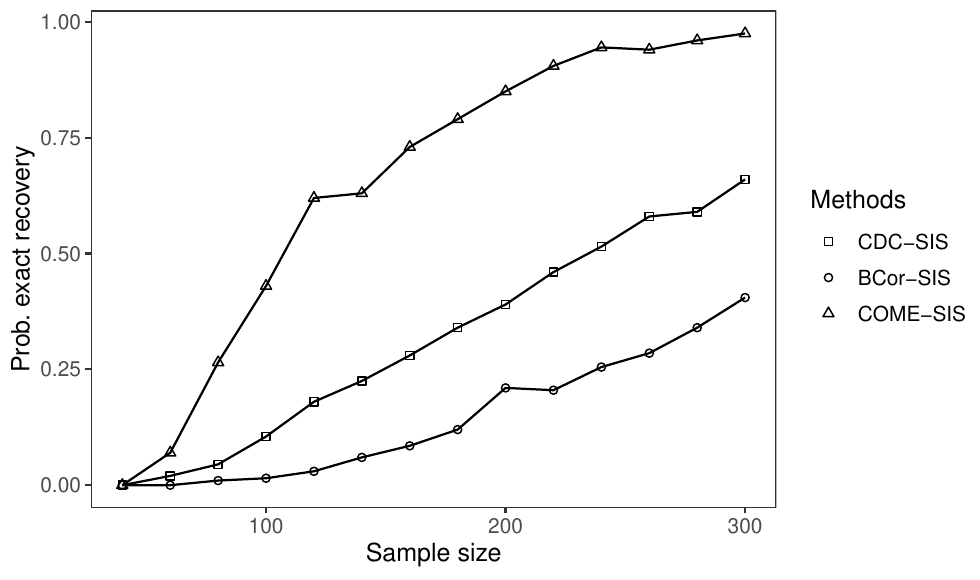}
  \end{center}
  \caption{Plot of the probability $\mathbb{P}(\hat{\mathcal{M}}_{d_n} = \mathcal{A})$ 
  versus the sample size under Model (1.b) with $d_n=3$,described in Section~\ref{sec:functional-simulation}.}
  \label{fig:strong-screening}
\end{figure}

Next, we study the strong screening property of the COME-CSIS,  
which ensures that $\hat{\mathcal{M}}_{d_n} = \mathcal{A}$ with high probability. 
The strong screening property asserts a more desirable result than the sure screening property, which only guarantees $\mathcal{A} \subseteq \hat{\mathcal{M}}_{d_n}$. 
Such a property is also for recently advanced screening methods such as \citet{huang2014feature} and \citet{pan2019generic}. 
As a demonstration, we apply the COME-CSIS method to synthetic datasets as described in Section~\ref{sec:functional-simulation}, setting $d_n = 3$. We compare this approach with BCor-SIS \citep{pan2019generic} and CDC-SIS \citep{wen2018sure}. The proportion of $\hat{\mathcal{M}}_{d_n} = \mathcal{A}$ is depicted in Figure~\ref{fig:strong-screening}. According to Figure~\ref{fig:strong-screening}, the likelihood that our proposed method precisely identifies set $\mathcal{A}$ approaches unity as the sample size increases. Concurrently, the performance of our method surpasses that of the methods compared.

To derive the strong screening of the COME-CSIS, 
we impose some conditions as follows. 
\begin{enumerate}[label=(C\arabic*), start=3]
  \item The kernel function $K(\cdot)$ is non-negative and uniformly-bounded, 
  and the bandwidth for kernel estimation of $Z$ satisfies $h=O(n^{-\kappa/{(2 d_z)}})$, 
  where $0 \leq \kappa < 1/2$.\label{kernel-condition}
  \item If $Z_1,\ldots,Z_6$ are independent copies of
  $Z$, then for $1\leq r\leq p$,
  there exists a positive constant $L$, such that
  \begin{align*}
\max_r |E(\zeta_{123456,r}|Z_1,Z_2,Z_3,Z_4,Z_5,Z_6)-E(\zeta_{123456,r}|Z'_1,Z_2,Z_3,Z_4,Z_5, Z_6)| \le L \| Z_1 - Z^\prime_1\|.
  \end{align*}\label{cbc-kernel-condition}
  \item There exists a positive constant $s_0$ such that for all $0<s\leq s_0$ and $z \in supp\{Z\}$,
  \begin{align*}
  E[\exp(s K_{H}(Z - z) \|Z - z\|)]<\infty.
  \end{align*}\label{subexponential-tail-distance}
  \item There exist some constants $c>0$ and $0\leq\kappa<1/2$ such that
  \begin{align*}
  \min_{r\in\mathcal{A}}\tau_r \geq 2cn^{-\kappa}.
  \end{align*}\label{larger-cbc}
\end{enumerate}
Condition~\ref{kernel-condition} imposes a mild condition on the kernel function, and 
most existing kernel functions satisfy this regularity condition. 
It also claims that the bandwidth of kernel estimation of $Z$ satisfies $h=O(n^{-\kappa/{(2 d_z)}})$. 
This requirement for bandwidth helps guarantee the density estimate is consistent. 
The works of \citet{liu2014feature} and \citet{wen2018sure} also rely on similar conditions. 
To make condition~\ref{kernel-condition} hold, the kernel function is set as a rectangle kernel, and its bandwidth is set as $h = n^{-1/(6d_z)}$ in our implementation.   
Condition~\ref{cbc-kernel-condition} is satisfied if the first order 
partial derivative of $E(\zeta_{123456,r}\mid Z_1, Z_2, Z_3, Z_4, Z_5, Z_6)$ are all bounded, 
which is similar to one of condition in \citet{wen2018sure}.
Condition~\ref{subexponential-tail-distance} provides an exponential bound on the product of kernel function and the distance of variable $Z$. 
Condition~\ref{larger-cbc} assumes that the minimum true signal has a lower bound with the order of $n^{-\kappa}$, 
which still allows the minimum true signal to approach zero when the sample size $n$ increases infinity. 
Assumptions analogous to condition~\ref{larger-cbc} are common in the high-dimensional data analysis literature 
(e.g., \citet{meinshausen2006ns}, \citet{fan2008sure}, \citet{barut2016csis}, \citet{liang2017robustscreening}).
Under these conditions, the following theorem presents the strong screening property of the COME-CSIS:

\begin{theorem}[Strong screening property of the COME-CSIS]\label{theorem:screening}
  If both the conditions in Theorem~\ref{theorem:equivalence} and conditions~\ref{kernel-condition}-\ref{subexponential-tail-distance} hold, 
  then for any $\tau_n \in (0, 2 c n^{\kappa})$, $0<\gamma<1/2-\kappa$, there exists positive constants $c_1$ and $c_2$ such that
  \begin{align*}
  \mathbb{P}(\max_{1\leq r\leq p}| \hat{\tau}_r-\tau_r| \geq
  cn^{-\kappa})\leq
  p[\exp(-c_1n^{1-2(\gamma+\kappa)})+n^6\exp(-c_2n^\gamma)]+o(1).
  \end{align*}
  If condition~\ref{larger-cbc} also holds, we have
  \begin{align*}
  \mathbb{P}(\mathcal {A}\subseteq \hat{\mathcal {M}}_{d_n}) &\geq
  1-|\mathcal{A}|\big[\exp(-c_1n^{1-2(\gamma+\kappa)})+n^6\exp(-c_2n^\gamma) + \exp(-c_3n^{1-2\kappa})\big]+o(1),\\
  \mathbb{P}(\hat{\mathcal{M}}_{d_n}\subseteq \mathcal {A}) &\geq
  1-d_n\big[\exp(-c_1n^{1-2(\gamma+\kappa)})+n^6\exp(-c_2n^\gamma) + \exp(-c_3n^{1-2\kappa})\big]+o(1), 
  \end{align*}
  where $|\mathcal {A}|$ is the size of the set $\mathcal {A}$ and $c_3$ is a positive constant.
  If $\log{p}=o(n^{\min\{1-2(\gamma + \kappa), \gamma\}})$, then the strong screening consistency holds, 
  \begin{align*}
    \mathbb{P}(\hat{\mathcal{M}}_{d_n} = \mathcal{A}) \stackrel{a.s.}{\longrightarrow} 1 \textup{ when } n \to \infty.
  \end{align*}
\end{theorem}

Theorem~\ref{theorem:screening} claims that, without specifying a model among $X_r, Y$, and $Z$, 
the COME-CSIS enjoys the strong screening property for the directionally limited metric space-valued response. 
Thus, it works for data in typical non-Euclidean spaces like separable Banach spaces, 
Riemannian manifolds, etc.
The COME-CSIS imposes no moment assumption on $X_1, \ldots, X_p$ and $Y$, and thus, it is robust to heavy-tailed observations and the presence of potential outliers appearing in these variables.  
This property is beneficial for high-dimensional data analysis because it is rigorous to assume all explanatory variables are well-behavior when $p$ is large.
Theorem~\ref{theorem:screening} also provides a theoretical guarantee for accurately selecting all relevant factors when analyzing non-Euclidean response variables in real applications.

As a complementary, we discuss determining the model size $d_n$. Generally speaking, 
it involves conducting a conditional independence test, 
a tough hypothesis test \citep{peter2020hardness}. 
Fortunately, if we assume the dependence between $Y$ and $(X, Z)$ follows one of the models: 
global Fr\'{e}chet regression \citep{muller2019frechet}, geodesic regression \citep{fletcher2013geodesic} or intrinsic regression \citep{cornea2016regression}, 
then we could stop adding a new predictor when it does not statistically affect the interpretation of the variation of $Y$. 
This effect can be measured by a generalization of the adjusted coefficient of determination for 
the global Fr\'{e}chet regression and geodesic regression, 
or a Wald-type test statistic for the intrinsic regression model. 

\section{Simulation}\label{sec:Emp}
In this section, we investigate the empirical performance of the COME-CSIS by synthesis datasets.
We evaluate the COME-CSIS by the proportion $P_i$ that it correctly includes the effective variable $X_i$ 
under a given model size $d_n$ in the $R$ replications \citep{Li2012}.
We also consider the proportion $P_{\textup{a}}$ that all effective variables are selected simultaneously. 
If the screening procedure works well, $P_i$s and $P_{\textup{a}}$ should be close to 1.
We fix the sample size $n$ at 150, the dimension $p$ at 2000, and the number of replications $R$ at 100.
We set the model size $d=\gamma \lceil n/\log(n)\rceil$ as \citet{Li2012} suggested, 
where $\lceil a \rceil$ refers to the integer part of $a$ 
and $\gamma$'s value is $1, 2,$ or $3$. 
For comparison, the BCor-SIS \citep{pan2019generic} and CDC-SIS \citep{wen2018sure} are taken into consideration, 
and two R packages: Ball and cdcsis \citep{zhu2018ball,hu2019cdcsis} are used to perform the two screening procedures. 
Since the CDC-SIS also relies on kernel density estimation, for fairness,  
the kernel function and its bandwidth used in the CDC-SIS 
is controlled to be the same as that of the COME-CSIS. 

We consider two common data types of non-Euclidean response: 
functional data and directional data, 
in Section~\ref{sec:functional-simulation} and 
Section~\ref{sec:directional-simulation}, respectively. 
As a complementary, the study of the classical Euclidean response is deferred to the supplementary material. 
For the conditional variable and explanatory variables $(Z, X_1, \ldots, X_p)$, 
they are jointly sampled from a multivariate normal distribution with zero means and 
covariance matrix $\Sigma$ where $\Sigma_{ij} = 2\rho^{|i - j|}$. 
The value of $\rho$ may be 0.0 or 0.7.


\subsection{Functional curve data}\label{sec:functional-simulation}
First, we entail the procedure for synthesizing the functional curve response. 
Suppose $B_1(t), \ldots, B_4(t)$ are four B-spline basis functions, we generate the a response $Y$ as the linear combination of the basis functions, where the linear coefficients are related to $Z, X_1, X_3, X_6$. We consider the following two models. 
The first model is a generalized additive model, 
and the second one is an interaction model:
\begin{align*}
  (1.a):  \quad  & Y(t) = Z B_1(t) + X_1 B_2(t) + X_3 B_3(t) + X_6^2 B_4(t) + \epsilon(t); \\
  (1.b):  \quad  & Y(t) = Z B_1(t) + 2 X_1 X_3 B_2(t) + Z X_3 B_3(t) + Z X_6 B_4(t) + \epsilon(t). 
\end{align*}
$\epsilon(t)$ is a zero-mean Gaussian process in these two models. 
The functional curve $Y(t)$ is observed in 17 equally spaced points at interval $[0, 8 \pi]$. 
To conduct the three screening procedures, 
$L_{2}$ norm \citep{manuel2012fda} are used to 
measure the dissimilarity of two functional observations. 

The results on functional data are displayed in Table \ref{tab:functional-simulation}. 
For Model (1.a), the three methods have a high probability of identifying the true variables, 
and both CDC-SIS and BCor-SIS are slightly better than the COME-CSIS. 
The CDC-SIS surpasses COME-CSIS because distance correlation-based methods have strong power in detecting the linear relationship \citep{pan2019ball}; 
BCor-SIS has better power because the effect of the conditional variable $Z$ can be considered additive noise. 
In Model (1.b), the COME-CSIS turns into the most powerful method and has a much larger chance to detect all influential variables simultaneously when $\rho = 0$. 
It is also worth noting that BCor-SIS has limited power to detect the influential variable $X_6$ that interacts with the conditional variable. 

\begin{table}[htbp]
  \centering
  \caption{The selection proportion of effective variables in functional data}
\begin{tabular}{c|c|c|cccc|cccc|}
  \toprule
  \multirow{2}{*}{$\rho$} & \multirow{2}{*}{$\gamma$} & \multirow{2}{*}{Methods} & \multicolumn{4}{c|}{Model (1.a)} & \multicolumn{4}{c|}{Model (1.b)} \\ 
  \cline{4-11}
  & & & $P_1$ & $P_3$ & $P_6$ & $P_a$ & $P_1$ & $P_3$ & $P_6$ & $P_a$ \\
  \midrule
  \multirow{9}{*}{0.0}  
  & $d_1$ & CDC-SIS   & 1.00 & 1.00 & 1.00 & 1.00 & 0.75 & 0.99 & 1.00 & 0.75 \\
  & $d_2$ & BCor-SIS  & 1.00 & 0.99 & 1.00 & 0.99 & 0.93 & 1.00 & 0.48 & 0.43 \\
  & $d_3$ & COME-CSIS  & 0.97 & 0.93 & 1.00 & 0.90 & 0.97 & 1.00 & 1.00 & 0.97 \\
  \cline{2-11}
  & $d_1$ & CDC-SIS   & 1.00 & 1.00 & 1.00 & 1.00 & 0.83 & 1.00 & 1.00 & 0.83 \\
  & $d_2$ & BCor-SIS  & 1.00 & 1.00 & 1.00 & 1.00 & 0.96 & 1.00 & 0.61 & 0.58 \\
  & $d_3$ & COME-CSIS  & 0.99 & 0.97 & 1.00 & 0.96 & 0.98 & 1.00 & 1.00 & 0.98 \\
  \cline{2-11}
  & $d_1$ & CDC-SIS   & 1.00 & 1.00 & 1.00 & 1.00 & 0.88 & 1.00 & 1.00 & 0.88 \\
  & $d_2$ & BCor-SIS  & 1.00 & 1.00 & 1.00 & 1.00 & 0.98 & 1.00 & 0.71 & 0.69 \\
  & $d_3$ & COME-CSIS  & 0.99 & 0.98 & 1.00 & 0.97 & 0.98 & 1.00 & 1.00 & 0.98 \\
  \midrule
  \multirow{9}{*}{0.7}  
  & $d_1$ & CDC-SIS   & 1.00 & 1.00 & 1.00 & 1.00 & 1.00 & 1.00 & 1.00 & 1.00 \\
  & $d_2$ & BCor-SIS  & 1.00 & 1.00 & 1.00 & 1.00 & 1.00 & 1.00 & 0.51 & 0.51 \\
  & $d_3$ & COME-CSIS  & 0.96 & 1.00 & 1.00 & 0.96 & 0.99 & 1.00 & 1.00 & 0.99 \\
  \cline{2-11}
  & $d_1$ & CDC-SIS   & 1.00 & 1.00 & 1.00 & 1.00 & 1.00 & 1.00 & 1.00 & 1.00 \\
  & $d_2$ & BCor-SIS  & 1.00 & 1.00 & 1.00 & 1.00 & 1.00 & 1.00 & 0.60 & 0.60 \\
  & $d_3$ & COME-CSIS  & 0.97 & 1.00 & 1.00 & 0.97 & 1.00 & 1.00 & 1.00 & 1.00 \\
  \cline{2-11}
  & $d_1$ & CDC-SIS   & 1.00 & 1.00 & 1.00 & 1.00 & 1.00 & 1.00 & 1.00 & 1.00 \\
  & $d_2$ & BCor-SIS  & 1.00 & 1.00 & 1.00 & 1.00 & 1.00 & 1.00 & 0.71 & 0.71 \\
  & $d_3$ & COME-CSIS  & 0.99 & 1.00 & 1.00 & 0.99 & 1.00 & 1.00 & 1.00 & 1.00 \\
  \bottomrule
\end{tabular}
\label{tab:functional-simulation}
\end{table}

\subsection{Directional data}\label{sec:directional-simulation}
We consider two models for generating directional data. 
The first model generates observations in a unit sphere, 
while the second generates observations in a unit circle. 
In the following, $U_c(a, b)$ represents a uniform distribution on the unit circle, 
where $a$ and $b$ are the lower and upper bound of the radian. 
The two models are: 
\begin{align*}
  (2.a):  \quad & \phi = Z X_3,~ \theta = \frac{1}{2}(X_6 + X_9) \\
  & Y = (\sin\theta\cos\phi, \cos\theta\sin\phi, \cos\theta). \\
  (2.b):  \quad & \textup{$\omega_1, \omega_2, \omega_3$ are $i.i.d$ Bernoulli random variables with $\frac{1}{2}$ probability to take value 1, }\\
  Y &=
  \begin{cases}
  \omega_1 U_c(0, \frac{\pi}{6}) + (1-\omega_1)U_c(\pi, \frac{7\pi}{6}), & \text{if $Z^\prime=1, X^\prime_3=I(X_3 > \texttt{median}(X_3))$} \\
  \omega_1 U_c(\frac{\pi}{2},\frac{2\pi}{3}) + (1-\omega_1)U_c(\frac{3\pi}{2}, \frac{5\pi}{3}), & \text{if $Z^\prime=1, X^\prime_3=I(X_3 \leq \texttt{median}(X_3))$ } \\
  \omega_2 U_c(\frac{\pi}{6},\frac{\pi}{3}) + (1-\omega_2)U_c(\frac{7\pi}{6},\frac{4\pi}{3}),  & \text{if $Z^\prime=2, X^\prime_6=I(X_6 > \texttt{median}(X_6))$} \\
  \omega_2 U_c(\frac{2\pi}{3},\frac{5\pi}{6}) + (1-\omega_2)U_c(\frac{5\pi}{3},\frac{11\pi}{6}), & \text{if $Z^\prime=2, X^\prime_6=I(X_6 \leq \texttt{median}(X_6))$ } \\
  \omega_3 U_c(\frac{\pi}{3},\frac{\pi}{2}) + (1-\omega_3)U_c(\frac{4\pi}{3},\frac{3\pi}{2}),  & \text{if $Z^\prime=3, X^\prime_9=I(X_9 > \texttt{median}(X_9))$} \\
  \omega_3 U_c(\frac{5\pi}{6},\pi) + (1-\omega_3)U_c(\frac{11\pi}{6},2\pi), & \text{if $Z^\prime=3, X^\prime_9=I(X_9 \leq \texttt{median}(X_9))$ }
  \end{cases}, \\
  &\textup{where $Z^\prime = I(Z \leq z') + 2 I(z' < Z \leq z'') + 3 I(z'' > Z)$, 
  and $z', z''$ are $\frac{1}{3}$ and}\\
  &\textup{$\frac{2}{3}$ quantiles of $Z$, respectively.}
\end{align*}
The geodesic distance measures the dissimilarity of two directional observations.

The performances of the three screening procedures on Models (2.a) and (2.b) are exhibited in Table~\ref{tab:spherical-simulation}.  
From the results of (2.a) in Table~\ref{tab:spherical-simulation}, 
all methods are powerful in identifying 
the variables $X_6, X_9$ that have no interaction with 
the conditional variable; 
however, only the COME-CSIS is good at detecting the effective variable $X_3$
that has an interaction effect on $Y$.  
Furthermore, under Model (2.b), the CDC-SIS cannot select all influential variables simultaneously  
since distance covariance works on metric space satisfying a strong negative type condition \citep{lyons2013distance}.
Nevertheless, the COME-CSIS and BCor-SIS still have powerful performance, generally surpassing the BCor-SIS.  

\begin{table}[htbp]
  \centering
  \caption{The selection proportion of effective variables in directional data}
\begin{tabular}{c|c|c|cccc|cccc|}
  \toprule
  \multirow{2}{*}{$\rho$} & \multirow{2}{*}{$\gamma$} & \multirow{2}{*}{Methods} & \multicolumn{4}{c|}{Model (2.a)} & \multicolumn{4}{c|}{Model (2.b)} \\ 
  \cline{4-11}
  & & & $P_3$ & $P_6$ & $P_9$ & $P_a$ & $P_3$ & $P_6$ & $P_9$ & $P_a$ \\
  \midrule
  \multirow{9}{*}{0.0}  
  & $d_1$ & CDC-SIS   & 0.64 & 0.99 & 0.99 & 0.62 & 0.04 & 0.00 & 0.00 & 0.00  \\ 
  & $d_2$ & BCor-SIS  & 0.06 & 1.00 & 1.00 & 0.06 & 0.39 & 0.31 & 0.35 & 0.04  \\ 
  & $d_3$ & COME-CSIS  & 0.96 & 0.92 & 0.96 & 0.85 & 0.76 & 0.70 & 0.69 & 0.36  \\ 
  \cline{2-11}
  & $d_1$ & CDC-SIS   & 0.77 & 0.99 & 1.00 & 0.76 & 0.05 & 0.01 & 0.01 & 0.00  \\ 
  & $d_2$ & BCor-SIS  & 0.08 & 1.00 & 1.00 & 0.08 & 0.59 & 0.48 & 0.54 & 0.18  \\ 
  & $d_3$ & COME-CSIS  & 0.98 & 0.95 & 0.98 & 0.91 & 0.92 & 0.84 & 0.83 & 0.64  \\ 
  \cline{2-11}
  & $d_1$ & CDC-SIS   & 0.83 & 1.00 & 1.00 & 0.83 & 0.07 & 0.01 & 0.01 & 0.00  \\ 
  & $d_2$ & BCor-SIS  & 0.11 & 1.00 & 1.00 & 0.11 & 0.73 & 0.59 & 0.65 & 0.30  \\ 
  & $d_3$ & COME-CSIS  & 0.99 & 0.97 & 0.99 & 0.95 & 0.94 & 0.90 & 0.92 & 0.78  \\ 
  \midrule
  \multirow{9}{*}{0.7}  
  & $d_1$ & CDC-SIS   & 0.61 & 1.00 & 1.00 & 0.61 & 0.01 & 0.01 & 0.01 & 0.00 \\
  & $d_2$ & BCor-SIS  & 0.08 & 1.00 & 1.00 & 0.08 & 0.55 & 0.69 & 0.65 & 0.25 \\
  & $d_3$ & COME-CSIS  & 0.93 & 1.00 & 0.99 & 0.93 & 0.54 & 0.79 & 0.81 & 0.37 \\
  \cline{2-11}
  & $d_1$ & CDC-SIS   & 0.76 & 1.00 & 1.00 & 0.76 & 0.01 & 0.03 & 0.02 & 0.00 \\
  & $d_2$ & BCor-SIS  & 0.15 & 1.00 & 1.00 & 0.15 & 0.75 & 0.87 & 0.83 & 0.55 \\
  & $d_3$ & COME-CSIS  & 0.96 & 1.00 & 1.00 & 0.96 & 0.71 & 0.91 & 0.90 & 0.59 \\
  \cline{2-11}
  & $d_1$ & CDC-SIS   & 0.80 & 1.00 & 1.00 & 0.80 & 0.02 & 0.05 & 0.03 & 0.00 \\
  & $d_2$ & BCor-SIS  & 0.20 & 1.00 & 1.00 & 0.20 & 0.86 & 0.92 & 0.91 & 0.72 \\
  & $d_3$ & COME-CSIS  & 0.96 & 1.00 & 1.00 & 0.96 & 0.79 & 0.96 & 0.92 & 0.71 \\
  \bottomrule
\end{tabular}
\label{tab:spherical-simulation}
\end{table}

\section{Screening of Corpus Callosum Shape}\label{sec:App}

Imaging genetics has gradually become an up-and-coming field.
Combining genetics and functional neuroimaging allows scientists to detect the association between neuroimaging phenotype and genetic variation.
In this section, we analyze databases with imaging, genetic, and clinical data provided by the Alzheimer's Disease Neuroimaging Initiative (ADNI) study\footnote{\url{adni.loni.usc.edu}} to demonstrate the helpfulness of the COME-CSIS.

\begin{figure}
	\centering
	\includegraphics[scale=0.8]{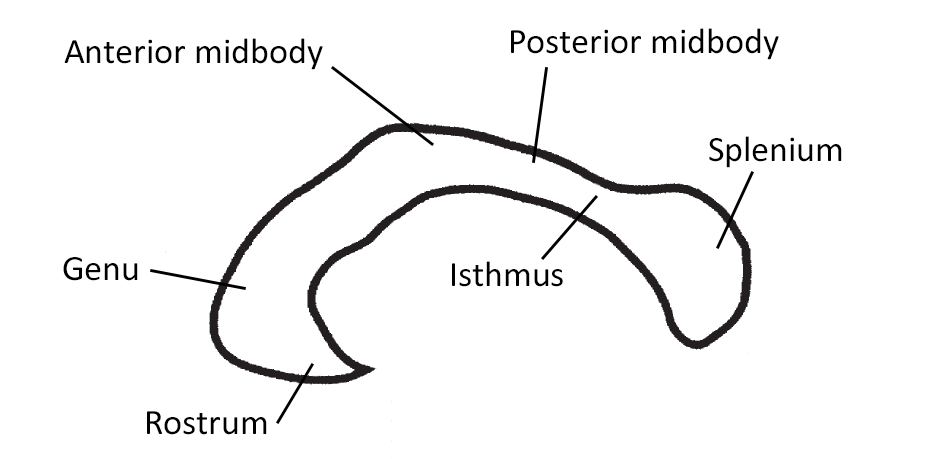}
	\caption{The image of midsagittal cut through the corpus callosum in \citet{eccher2014cc} with its major subunits from anterior to posterior: rostrum, genu, anterior midbody, posterior midbody, isthmus, and splenium.}
    \label{fig:anatonmy_corpus_callosum}
\end{figure}

The corpus callosum is the largest white matter structure in the human brain,
connecting the left and right hemispheres in the brain \citep{edwards2014cc}.
Anatomically, CC is roughly a 10 cm long, 1 cm wide curved structure in an adult human, containing approximately 200 million fibers.
Standard anatomic divisions for CC are the rostrum, genu, body (including anterior and posterior parts),
isthmus, and splenium (see Figure~\ref{fig:anatonmy_corpus_callosum}).
Functionally, CC is essential for communication between the two cerebral hemispheres.
It facilitates the integration of sensory and motor information from the two sides of the body
and different influences on higher cognition related to social interaction and language.
Since CC is an essential structural and functional part of the brain, finding the factors that impact its shape is valuable.
Existing literature has shown that age and
disease are related to CC shape \citep{hinkle2014intrinsic,cornea2016regression, fletcher2013geodesic, joshi2013statistical};
However, the genetic factors influencing the CC shape have been largely undiscovered.

To this end, we analyze a preprocessed ADNI dataset \citep{cornea2016regression} 
containing CC shapes and demographics. 
The preprocessed dataset includes 371 subjects, 
where each subject's gender, age, disease diagnosis (AD or health), 
CC's shape and 6,995 single nucleotide polymorphisms (SNPs) are recorded.
50 two-dimensional points characterize the CC shape of each subject 
on the CC contour. Moreover, the dissimilarity between 
two shape observations are measured by Riemannian shape distance \citep{dryden2016statistical}. 

Next, we specify the explanatory variables and conditional variables in our analysis. 
Due to the difference in shape along the inner side of the posterior splenium and isthmus subregions for different age groups, 
it is concluded that age strongly influences the shape of CC contours \citep{cornea2016regression}.
Besides, \citet{cornea2016regression} points out that the splenium seems less rounded, and the isthmus is thinner in subjects with AD than in HC. 
We conducted additional tests to assess the correlation between the CC shape and three variables: age, gender, and disease status. These tests were performed using the ball covariance method with permutation replications $499$ \citep{pan2019ball}. The results are summarized in Table \ref{tab:covtest}.
\begin{table}[h]
    \centering
    \caption{P-values from correlation tests between CC shape and age, gender, and disease status}
    \begin{tabular}{c|c|c|c}
        \toprule
         & Age  & Gender & Disease status\\
         \midrule
      p-value   & 0.01  & 0.09 & 0.02 \\
      \bottomrule
    \end{tabular} \label{tab:covtest}
\end{table}

At a significance level of 0.05, both age and disease status showed significant correlations with CC shape. In contrast, the correlation between
gender and CC shape was not statistically significant.
Given the two prior information, our interest is to detect the genetic factors associated with CC contour when controlling covariates age and group.
Moreover, to verify the arguments that there is no significant gender effect on the CC shape after disease status and age are considered \citep{cornea2016regression}, 
we treat gender as one of the explanatory variables in the screening procedure. We set $d_n = n / \log(n)$ and select the previously identified important SNPs for further analysis.


Table~\ref{tab:cc-gene} lists the top 5 important SNPs 
selected for CC shape by the BCor-SIS, CDC-SIS, and COME-CSIS.
From Table~\ref{tab:cc-gene}, 
the COME-CSIS and CDC-SIS select similar top-5 SNPs, 
but BCor-SIS selects quite different SNPs, 
This implies that considering the conditional variable greatly impacts the screening results.
The SNP rs11668269, selected by the COME-CSIS and CDC-SIS, 
is in the LOC105372330 gene that is negatively correlated to APOE4+ males frontal white matter, 
where the APOE4 allele causes damage to the white matter of the corpus callosum \citep{hsu2019sexdifference, koizumi2018apovarepsilon4}.

The SNP rs4807022, initially identified through BCor-SIS, is within the gene encoding protein tyrosine phosphatase receptor sigma (PTPRS). This placement suggests a significant association with the CC shape. We conducted a hypothesis test on SNP rs4807022 with the conditional variables to further explore this relationship, employing ball covariance. The analysis yielded a p-value of 0.0267 for the association between the SNP and age, indicating a strong correlation. Conversely, the p-value for the association between the SNP and Alzheimer’s Disease (AD) was 0.99, suggesting no significant dependence. We then segmented the data into age groups based on quantiles (55--72, 73--76, 77--80, and 81--95\footnote{55, 72, 76, 80, and 92 are 0.0, 0.25, 0.5, 0.75, and 1.0 quantiles of the age, respectively.}) to examine variations across different age segments. This SNP data categorizes genetic variations as 0 (AA), 1 (Aa), and 2 (aa), representing two, one, and zero copies of the reference allele, respectively. The results, summarized in Table~\ref{tab:cc-snp1}, reveal that the frequency of heterozygotes (1 for genotype Aa) increases with age. 
PTPRS has been implicated in the regulation of neurite outgrowth and has been shown to stimulate neurite outgrowth in response to the heparan sulfate proteoglycan GPC2, essential for normal brain development \citep{pulido1995lar,coles2011proteoglycan}. These functions suggest that PTPRS may play a role in longevity.
Despite these findings, SNP rs4807022 is ranked 4808th in COME-CSIS, and when conditioned on age, the p-value derived from the COME analysis is 0.37.
Overall, while there is a strong association between SNP rs4807022 and age, its link to CC shape appears to be mediated primarily through age-related changes rather than a direct genetic influence on CC shape.

Moreover, the COME-CSIS selects the SNP rs3745129 in the ZNF329 gene as the fourth important gene. 
Because an important paralog of the ZNF329 gene, ZFP37, 
expresses in all neurons of the central nervous system \citep{mazarakis1996zfp37}, 
ZNF329 may influence the corpus callosum. 
Finally, the gender factor is ranked 4553 by the COME-CSIS, 
implying the gender effect is not significant after considering the age and disease status. 
This result is coincident with the finding of \citet{cornea2016regression}.

To further compare the performance of COME-CSIS and CDC-SIS, we randomly generated 1,000 noise SNP variables and identified the proportion of noise selected by the two methods. 
We begin by considering the Major Allele Frequency ($p$) and Minor Allele Frequency (MAF, $q$) in SNP noise, $p + q = 1$. We then generate the frequencies of the three genotypes (AA, Aa, aa) according to the Hardy-Weinberg Equilibrium (HWE). We conduct $\chi^2$ test to check HWE, $5.86 \%$ of SNPs are rejected under significance level $0.05$, which indicates that the SNPs in our dataset predominantly adhere to HWE, so it is reasonable to generate noise SNPs based on HWE. To simulate SNP noise, we employed two methods. The first method involved calculating the average MAF of 6,994 SNPs, found to be 45.9\%, and subsequently generating 1,000 SNPs with MAF 45.9\% to serve as noise. The second method consisted of randomly selecting 1,000 SNPs from the original set of 6,994 and generating noise based on their actual MAF. We conducted this experiment 20 times to ensure statistical reliability and counted the proportion of SNPs included in the noise under different settings of $d_n = \gamma \lceil n/\log(n)\rceil, \gamma = 1, 2, 3$. 
The results in Table~\ref{tab:cc-snp2} demonstrate that the COME-CSIS method selects a higher proportion of original, non-noise SNPs than CDC-SIS. This indicates that COME-CSIS is more robust and less affected by noise in the data.
The superior performance of COME-CSIS is attributed to its utilization of rank-based distance metrics and local neighborhood structures, which significantly enhance its robustness against outliers and heavy-tailed distributions.

\begin{table}[!h]
\center
 \caption{Top-5 selected variables for corpus callosum shape by different methods. SNP's corresponding genes are in parentheses if they exist}\label{tab:cc-gene}
\begin{tabular}{ccc}
\toprule
  BCor-SIS & CDC-SIS & COME-CSIS \\
  \midrule
  rs4807022 (PTPRS) & rs10414182  & rs11668269 (LOC105372330) \\
  rs8101539 (WDR88)  & rs10424248 & rs10414182  \\
  rs3786627 (ETFB) & rs11668269 (LOC105372330) & rs3786776 (PLA2G4C)  \\ 
  rs11085313 & rs1979260 (MYO9B)  & rs3745129 (ZNF329)  \\ 
  rs4805863 & rs3786776 (PLA2G4C)   & rs10518253 \\
\bottomrule
\end{tabular}
\end{table}

\begin{table}[!h]
\center
 \caption{SNP value in four age groups.}\label{tab:cc-snp1}
\begin{tabular}{c|cccc}
\toprule
 & 55--72 & 73--76 & 77--80 & 81--95 \\
 \midrule
 AA & 70 & 72 & 65 & 50 \\
 \hline
  Aa & 35 & 13 & 25 & 33 \\
  \hline
  aa & 1 & 2 & 2 & 2 \\
\bottomrule
\end{tabular}
\end{table}

\begin{table}[!h]
\center
 \caption{The average proportion of SNP noise selected by the two methods}\label{tab:cc-snp2}
\begin{tabular}{ccccc}
\toprule
Sampling mechanism & Method & 63 ($\lceil n/\log(n)\rceil$) & 126 ($2\lceil n/\log(n)\rceil$) & 189 ($3\lceil n/\log(n)\rceil$) \\
 \midrule
\multirow{2}{*}{Random}& COME-CSIS & 7.94\% & 9.60\% & 11.06\% \\
 &  CDC-SIS & 13.02\% & 13.57\% & 12.70\%  \\
\hline
\multirow{2}{*}{Average}& COME-CSIS & 3.49\% & 7.14\% & 7.57\% \\
 & CDC-SIS & 8.41\% & 8.89\% & 8.41\%  \\
\bottomrule
\end{tabular}
\end{table}

\begin{figure}
\begin{center}
\includegraphics[scale=1.0]{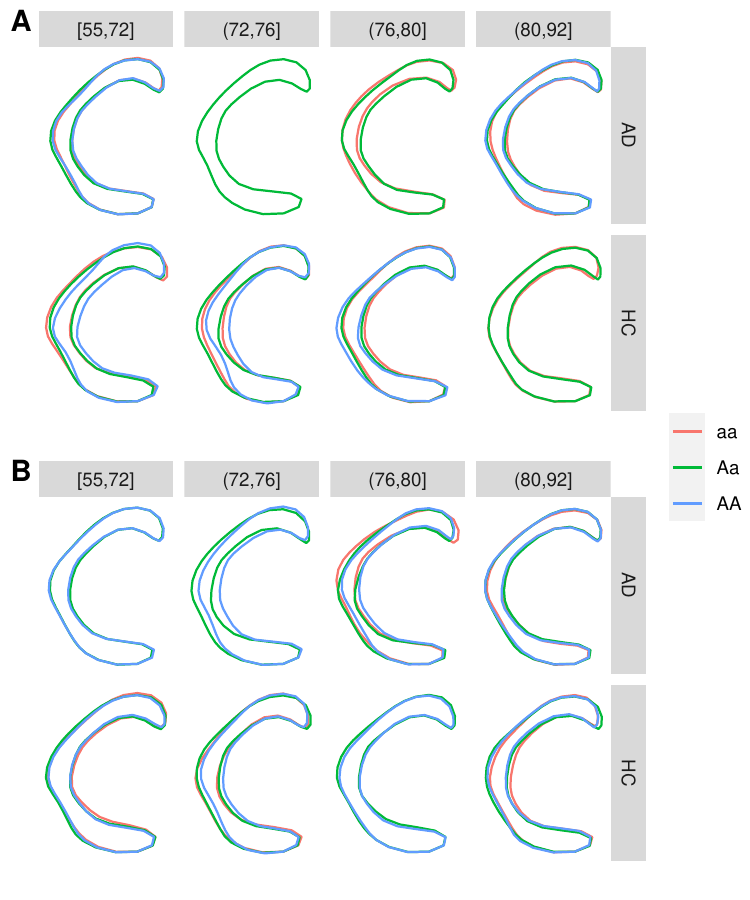}
\end{center}
\vspace{-15pt}
\caption{The corpus callosum (CC) shape in SNP rs11668269 (subfigure A) and rs10410302 (subfigure B).
In each subfigure, the lower panels indicate healthy control, the upper panels indicate patients with Alzheimer's disease, 
and a column represents a specific age group. Four columns represent 55--72, 73--76, 77--80, and 81--95 years old, respectively. 
Ordinary Procrustes analysis \citep{dryden2016statistical} are conducted for all CC shapes. 
}
\label{fig:corpus_callosum}
\end{figure}

We visualize CC shapes in Figure~\ref{fig:corpus_callosum} to gain more insight. 
Specifically, we divide the subjects into eight disjoint subsets according to their ages phase (55--72, 73--76, 77--80, and 81--95) 
and disease status (AD and HC), 
then, in each part of the subject, we demonstrate the Fr\'{e}chet mean of CC shapes when the genetic factor takes different values. 
It can be observed that, for both of the top-2 SNPs selected by the COME-CSIS, 
they impact the CC shape condition on age and health status. 
More precisely, for the healthy subjects with age 72--76, 
the SNP rs11668269 with value AA associates with a smaller curvature in the body part of CC; 
for the AD patients with age 72-76, 
the SNP rs10410302 with value AA leads to a smaller curvature in the body part of CC.

In practical applications, COME-CSIS demonstrates enhanced capability in identifying key variables compared to methods that do not utilize conditional variables. By preemptively controlling for confounding factors such as age and gender, COME-CSIS allows for more accurate isolation and identification of core variables. 
Compared with other conditional variable screening methods (CDC-SIS), the COME-CSIS is more robust and less likely to select irrelevant noise.
Consequently, this approach improves the robustness and reliability of the findings in complex analytical scenarios where multiple interdependent variables are present.



\section{Conclusion and Discussion}\label{sec:dis}
We propose a conditional dependency measure named COME for metric spaces, which is capable of assessing both local and global dependencies. This model-free measure is adaptable to various complex non-Euclidean spaces. Although our initial analysis assumes $X$ to be Euclidean, we can extend $X$ to general metric space, provided that it satisfies the condition of being directionally limited. Additionally, we introduce a conditional screening method based on COME for ultra-high-dimensional space, proving its strong screening properties and offering theoretical support for practical applications.

Our method was employed to analyze genes associated with the shape of the corpus callosum. After integrating insights from both literature and data analysis, we utilized Alzheimer's Disease status and age as conditional variables $Z$, with CC shape as the response variable $Y$ and gene as the predictor variable $X$. Most existing methods falter when applied to non-Euclidean response variables like CC shape. We identified the five most relevant SNPs for further analysis. The scientific validity of our SNP selection has been corroborated by existing research, which highlights connections to white matter and the APOE4+ allele. By contrasting our approach with non-conditional screening techniques like BCor-SIS, we demonstrated that while the selected SNPs show strong correlations with factors such as age, they are not directly pivotal in influencing CC shape. Moreover, we simulated noisy SNPs based on the Hardy-Weinberg Equilibrium criterion, thereby demonstrating the robustness of COME-SIS and its capacity to mitigate the impact of irrelevant SNPs. Overall, in the context of CC shape-related gene analysis, COME-SIS proved effective and resilient in handling complex non-Euclidean response variables and high-dimensional predictor variables, showcasing broad potential for future applications.

Several issues warrant further investigation. The threshold used in our method is adopted from \citep{fan2008sure}. It would be of interest to develop a criterion to determine the threshold for samples, which we leave as a subject for future research. Moreover, when processing the response variable $Z$ through kernel technology, we encounter the curse of dimensionality, particularly when $Z$ is high-dimensional. This scenario is not uncommon, particularly when conditioning on a set of genes. Thus, developing advanced conditional analysis techniques to circumvent the curse of dimensionality represents a crucial direction for future research. Lastly, while we have successfully identified key SNPs in our real data analysis, it is also worthwhile to explore further analyses, such as predicting the shape of CC itself or developing more effective prognostic strategies.

\begin{acks}[Acknowledgments]
    The first two authors contributed equally and are listed in alphabetical order.
    
    Dr.~Wang's research is partially supported by the National Natural Science Foundation of China (12231017, 72171216) and the National Key R\&D Program of China (2022YFA1003803).
  Dr.~Pan's research is partially supported by the National Natural Science Foundation of China (12322113, 72495122, 12288201). 

  
  Data collection and sharing for this project was funded by the Alzheimer's Disease Neuroimaging Initiative
(ADNI) (National Institutes of Health Grant U01 AG024904) and DOD ADNI (Department of Defense award
number W81XWH-12-2-0012). ADNI is funded by the National Institute on Aging, the National Institute of
Biomedical Imaging and Bioengineering, and through generous contributions from the following: AbbVie,
Alzheimer’s Association; Alzheimer’s Drug Discovery Foundation; Araclon Biotech; BioClinica, Inc.; Biogen;
Bristol-Myers Squibb Company; CereSpir, Inc.; Cogstate; Eisai Inc.; Elan Pharmaceuticals, Inc.; Eli Lilly and
Company; EuroImmun; F. Hoffmann-La Roche Ltd and its affiliated company Genentech, Inc.; Fujirebio; GE
Healthcare; IXICO Ltd.; Janssen Alzheimer Immunotherapy Research \& Development, LLC.; Johnson \&
Johnson Pharmaceutical Research \& Development LLC.; Lumosity; Lundbeck; Merck \& Co., Inc.; Meso
Scale Diagnostics, LLC.; NeuroRx Research; Neurotrack Technologies; Novartis Pharmaceuticals
Corporation; Pfizer Inc.; Piramal Imaging; Servier; Takeda Pharmaceutical Company; and Transition
Therapeutics. The Canadian Institutes of Health Research is providing funds to support ADNI clinical sites
in Canada. Private sector contributions are facilitated by the Foundation for the National Institutes of Health
(www.fnih.org). The grantee organization is the Northern California Institute for Research and Education,
and the study is coordinated by the Alzheimer’s Therapeutic Research Institute at the University of Southern
California. ADNI data are disseminated by the Laboratory for Neuro Imaging at the University of Southern
California.

\end{acks}


\begin{supplement}
\stitle{Supplement A: Supplement for “Identification of Genetic Factors Associated with Corpus Callosum Morphology: Conditional Strong Independence Screening for Non-Euclidean Responses”}
\sdescription{The Supplementary Material includes technical proof and the detail of the iterative COME-CSIS algorithm.}
\end{supplement}

\begin{supplement}
    \stitle{Supplement B: R code for “Identification of Genetic Factors Associated with Corpus Callosum Morphology: Conditional Strong Independence Screening for Non-Euclidean Responses”}
    \sdescription{R-package \textit{come} containing code to perform the COME-CSIS described in the article and R scripts used to generate the numerical results presented in the paper.}
\end{supplement}


\bibliographystyle{imsart-nameyear} 
\bibliography{reference}       

@article{mazarakis1996zfp37,
  title = {Zfp-37Is a Member of the KRAB Zinc Finger Gene Family and Is Expressed in Neurons of the Developing and Adult CNS},
  journal = {Genomics},
  volume = {33},
  number = {2},
  pages = {247-257},
  year = {1996},
  issn = {0888-7543},
  doi = {https://doi.org/10.1006/geno.1996.0189},
  author = {N. Mazarakis and D. Michalovich and A. Karis and F. Grosveld and N. Galjart}
}

@book{federer2014geometric,
  title={Geometric measure theory},
  author={Federer, Herbert},
  year={2014},
  publisher={Springer}
}

@article{koizumi2018apovarepsilon4,
  title={Apo$\varepsilon$4 disrupts neurovascular regulation and undermines white matter integrity and cognitive function},
  author={Koizumi, Kenzo and Hattori, Yorito and Ahn, Sung Ji and Buendia, Izaskun and Ciacciarelli, Antonio and Uekawa, Ken and Wang, Gang and Hiller, Abigail and Zhao, Lingzhi and Voss, Henning U and others},
  journal={Nature communications},
  volume={9},
  number={1},
  pages={1--11},
  year={2018},
  publisher={Nature Publishing Group}
}

@article{liang2017robustscreening,
    author = {Jingnan Xue and Faming Liang},
    title = {A Robust Model-Free Feature Screening Method for Ultrahigh-Dimensional Data},
    journal = {Journal of Computational and Graphical Statistics},
    volume = {26},
    number = {4},
    pages = {803-813},
    year  = {2017},
    publisher = {Taylor \& Francis},
    doi = {10.1080/10618600.2017.1328364},
    note ={PMID: 30532512}
}

@article{wang2021nonparametric,
author = {Xueqin Wang and Jin Zhu and Wenliang Pan and Junhao Zhu and Heping Zhang},
title = {Nonparametric Statistical Inference via Metric Distribution Function in Metric Spaces},
journal = {Journal of the American Statistical Association},
volume = {119},
number = {548},
pages = {2772--2784},
year = {2024},
publisher = {ASA Website},
doi = {10.1080/01621459.2023.2277417},
URL = {https://doi.org/10.1080/01621459.2023.2277417},
eprint = {https://doi.org/10.1080/01621459.2023.2277417}
}

@article{muller2019frechet,
  author = "Petersen, Alexander and Müller, Hans-Georg",
  doi = "10.1214/17-AOS1624",
  fjournal = "Annals of Statistics",
  journal = "Ann. Statist.",
  month = "04",
  number = "2",
  pages = "691--719",
  publisher = "The Institute of Mathematical Statistics",
  title = "Fréchet regression for random objects with Euclidean predictors",
  url = "https://doi.org/10.1214/17-AOS1624",
  volume = "47",
  year = "2019"
}

@article{peter2020hardness,
  author = "Shah, Rajen D. and Peters, Jonas",
  doi = "10.1214/19-AOS1857",
  journal = "Annals of Statistics",
  ajournal = "Ann. Statist.",
  month = "06",
  number = "3",
  pages = "1514--1538",
  publisher = "The Institute of Mathematical Statistics",
  title = "The hardness of conditional independence testing and the generalised covariance measure",
  volume = "48",
  year = "2020"
}

@Article{hsu2019sexdifference,
  AUTHOR = {Hsu, M and Dedhia, M and Crusio, WE and Delprato, A},
  TITLE = {Sex differences in gene expression patterns associated with the APOE4 allele},
  JOURNAL = {F1000Research},
  VOLUME = {8},
  YEAR = {2019},
  NUMBER = {387},
  DOI = {10.12688/f1000research.18671.2}
}

@book{dryden2016statistical,
  title     = {Statistical shape analysis: with applications in R},
  author    = {Dryden, Ian L and Mardia, Kanti V},
  volume    = {995},
  year      = {2016},
  publisher = {John Wiley \& Sons},
  address   = {}
}

@incollection{eccher2014cc,
  title     = {Corpus Callosum},
  editor    = {Michael J. Aminoff and Robert B. Daroff},
  booktitle = {Encyclopedia of the Neurological Sciences (Second Edition)},
  publisher = {Academic Press},
  edition   = {Second Edition},
  address   = {Oxford},
  pages     = {867 - 868},
  year      = {2014},
  isbn      = {978-0-12-385158-1},
  doi       = {https://doi.org/10.1016/B978-0-12-385157-4.01137-4},
  author    = {M. Eccher}
}

@article{hinkle2014intrinsic,
  title     = {Intrinsic polynomials for regression on Riemannian manifolds},
  author    = {Hinkle, Jacob and Fletcher, P Thomas and Joshi, Sarang},
  journal   = {Journal of Mathematical Imaging and Vision},
  volume    = {50},
  number    = {1-2},
  pages     = {32--52},
  year      = {2014},
  publisher = {Springer}
}

@article{joshi2013statistical,
  title     = {Statistical shape analysis of the corpus callosum in schizophrenia},
  author    = {Joshi, Shantanu H and Narr, Katherine L and Philips, Owen R and Nuechterlein, Keith H and Asarnow, Robert F and Toga, Arthur W and Woods, Roger P},
  journal   = {Neuroimage},
  volume    = {64},
  pages     = {547--559},
  year      = {2013},
  publisher = {Elsevier}
}

@article{fletcher2013geodesic,
  title     = {Geodesic regression and the theory of least squares on Riemannian manifolds},
  author    = {Fletcher, P Thomas},
  journal   = {International journal of computer vision},
  volume    = {105},
  number    = {2},
  pages     = {171--185},
  year      = {2013},
  publisher = {Springer}
}

@article{huang2014feature,
  author = {Danyang Huang and Runze Li and Hansheng Wang},
  title = {Feature Screening for Ultrahigh Dimensional Categorical Data With Applications},
  journal = {Journal of Business \& Economic Statistics},
  volume = {32},
  number = {2},
  pages = {237-244},
  year  = {2014},
  publisher = {Taylor \& Francis},
  doi = {10.1080/07350015.2013.863158}
}

@article{meinshausen2006ns,
  author = "Meinshausen, Nicolai and Buhlmann, Peter",
  doi = "10.1214/009053606000000281",
  fjournal = "Annals of Statistics",
  journal = "Ann. Statist.",
  month = "06",
  number = "3",
  pages = "1436--1462",
  publisher = "The Institute of Mathematical Statistics",
  title = "High-dimensional graphs and variable selection with the Lasso",
  volume = "34",
  year = "2006"
}

@article{biswal2010toward,
	Author = {Biswal, Bharat B. and Mennes, Maarten and Zuo, Xi-Nian and Gohel, Suril and Kelly, Clare and Smith, Steve M. and Beckmann, Christian F. and Adelstein, Jonathan S. and Buckner, Randy L. and Colcombe, Stan and Dogonowski, Anne-Marie and Ernst, Monique and Fair, Damien and Hampson, Michelle and Hoptman, Matthew J. and Hyde, James S. and Kiviniemi, Vesa J. and K{\"o}tter, Rolf and Li, Shi-Jiang and Lin, Ching-Po and Lowe, Mark J. and Mackay, Clare and Madden, David J. and Madsen, Kristoffer H. and Margulies, Daniel S. and Mayberg, Helen S. and McMahon, Katie and Monk, Christopher S. and Mostofsky, Stewart H. and Nagel, Bonnie J. and Pekar, James J. and Peltier, Scott J. and Petersen, Steven E. and Riedl, Valentin and Rombouts, Serge A. R. B. and Rypma, Bart and Schlaggar, Bradley L. and Schmidt, Sein and Seidler, Rachael D. and Siegle, Greg J. and Sorg, Christian and Teng, Gao-Jun and Veijola, Juha and Villringer, Arno and Walter, Martin and Wang, Lihong and Weng, Xu-Chu and Whitfield-Gabrieli, Susan and Williamson, Peter and Windischberger, Christian and Zang, Yu-Feng and Zhang, Hong-Ying and Castellanos, F. Xavier and Milham, Michael P.},
	Doi = {10.1073/pnas.0911855107},
	Issn = {0027-8424},
	Journal = {Proceedings of the National Academy of Sciences},
	Number = {10},
	Pages = {4734--4739},
	Publisher = {National Academy of Sciences},
	Title = {Toward discovery science of human brain function},
	Volume = {107},
	Year = {2010}
}

@article{manuel2012fda,
   author = {Manuel Febrero-Bande and Manuel de la Fuente},
   title = {Statistical Computing in Functional Data Analysis: The R Package fda.usc},
   journal = {Journal of Statistical Software, Articles},
   volume = {51},
   number = {4},
   year = {2012},
   issn = {1548-7660},
   pages = {1--28},
   doi = {10.18637/jss.v051.i04},
}

@article{lin2015adaptive,
  title = "Adaptive conditional feature screening",
  journal = "Computational Statistics \& Data Analysis",
  volume = "94",
  pages = "287 - 301",
  year = "2016",
  issn = "0167-9473",
  doi = "10.1016/j.csda.2015.09.002",
  author = "Lu Lin and Jing Sun"
}

@article{hong2018conditional,
  title={Conditional screening for ultra-high dimensional covariates with survival outcomes},
  author={Hong, Hyokyoung G and Kang, Jian and Li, Yi},
  journal={Lifetime data analysis},
  volume={24},
  number={1},
  pages={45--71},
  year={2018},
  publisher={Springer}
}

@article{hu2017conditional,
  title={Conditional sure independence screening by conditional marginal empirical likelihood},
  author={Hu, Qinqin and Lin, Lu},
  journal={Annals of the Institute of Statistical Mathematics},
  volume={69},
  number={1},
  pages={63--96},
  year={2017},
  publisher={Springer}
}

@article{barut2016csis,
  author = {Emre Barut and Jianqing Fan and Anneleen Verhasselt},
  title = {Conditional Sure Independence Screening},
  journal = {Journal of the American Statistical Association},
  volume = {111},
  number = {515},
  pages = {1266-1277},
  year  = {2016},
  publisher = {Taylor \& Francis},
  doi = {10.1080/01621459.2015.1092974},
  note ={PMID: 28360436}
}

@Manual{hu2019cdcsis,
  title = {cdcsis: Conditional Distance Correlation Based Feature Screening and Conditional Independence Inference},
  author = {Wenhao Hu and Mian Huang and Wenliang Pan and Xueqin Wang and Canhong Wen and Yuan Tian and Heping Zhang and Jin Zhu},
  year = {2019},
  note = {R package version 2.0.3},
  url = {https://github.com/Mamba413/cdcsis},
}

@article{zhu2018ball,
  author = {Jin Zhu and Wenliang Pan and Wei Zheng and Xueqin Wang},
  title = {Ball: An R Package for Detecting Distribution Difference and Association in Metric Spaces},
  journal = {Journal of Statistical Software, Articles},
  volume = {97},
  number = {6},
  year = {2021},
  issn = {1548-7660},
  pages = {1--31},
  doi = {10.18637/jss.v097.i06},
  url = {https://www.jstatsoft.org/v097/i06}
}

@article{fan2008sure,
  title = {Sure independence screening for ultrahigh dimensional feature space},
  author = {Fan, Jianqing and Lv, Jinchi},
  journal = {Journal of the Royal Statistical Society: Series B (Statistical Methodology)},
  volume = {70},
  number = {5},
  pages = {849-911},
  doi = {10.1111/j.1467-9868.2008.00674.x},
  year = {2008}
}

@article{wang2015conditional,
  author = {Xueqin Wang and Wenliang Pan and Wenhao Hu and Yuan Tian and Heping Zhang},
  title = {Conditional Distance Correlation},
  journal = {Journal of the American Statistical Association},
  volume = {110},
  number = {512},
  pages = {1726-1734},
  year  = {2015},
  publisher = {Taylor \& Francis},
  doi = {10.1080/01621459.2014.993081},
  note ={PMID: 26877569}
}

@Article{lyons2013distance,
  author    = {Lyons, Russell},
  title     = {Distance Covariance in Metric Spaces},
  journal   = {The Annals of Probability},
  year      = {2013},
  volume    = {41},
  number    = {5},
  pages     = {3284--3305},
  month     = {09},
  ajournal  = {Ann. Probab.},
  doi       = {10.1214/12-AOP803},
  publisher = {The Institute of Mathematical Statistics},
}

@inproceedings{fukumizu2008kernel,
  title={Kernel measures of conditional dependence},
  author={Fukumizu, Kenji and Gretton, Arthur and Sun, Xiaohai and Sch{\"o}lkopf, Bernhard},
  booktitle={Advances in neural information processing systems},
  pages={489--496},
  year={2008}
}

@Article{pan2019ball,
  author    = {Wenliang Pan and Xueqin Wang and Heping Zhang and Hongtu Zhu and Jin Zhu},
  title     = {Ball Covariance: A Generic Measure of Dependence in Banach Space},
  journal   = {Journal of the American Statistical Association},
  volume = {115},
  number = {529},
  pages = {307-317},
  year  = {2020},
  doi       = {10.1080/01621459.2018.1543600},
  publisher = {Taylor \& Francis},
}

@article{Li2012,
author = { Runze   Li  and  Wei   Zhong  and  Liping   Zhu },
title = {Feature Screening via Distance Correlation Learning},
journal = {Journal of the American Statistical Association},
volume = {107},
number = {499},
pages = {1129-1139},
year  = {2012},
publisher = {Taylor \& Francis},
doi = {10.1080/01621459.2012.695654},
    note ={PMID: 25249709},
URL = { 
        https://doi.org/10.1080/01621459.2012.695654
},
eprint = { 
        https://doi.org/10.1080/01621459.2012.695654 
}
}

@article{edwards2014cc,
    author = {Edwards, Timothy J. and Sherr, Elliott H. and Barkovich, A. James and Richards, Linda J.},
    title = "{Clinical, genetic and imaging findings identify new causes for corpus callosum development syndromes}",
    journal = {Brain},
    volume = {137},
    number = {6},
    pages = {1579-1613},
    year = {2014},
    month = {01},
    issn = {0006-8950},
    doi = {10.1093/brain/awt358},
    url = {https://doi.org/10.1093/brain/awt358}
}

@article{cornea2016regression,
  author = {Cornea, Emil and Zhu, Hongtu and Kim, Peter and Ibrahim, Joseph},
  year = {2016},
  month = {03},
  pages = {n/a-n/a},
  title = {Regression Models on Riemannian Symmetric Spaces},
  volume = {79},
  journal = {Journal of the Royal Statistical Society: Series B (Statistical Methodology)},
  doi = {10.1111/rssb.12169}
}

@article{pan2019generic,
  title={A generic sure independence screening procedure},
  author={Pan, Wenliang and Wang, Xueqin and Xiao, Weinan and Zhu, Hongtu},
  journal={Journal of the American Statistical Association},
  volume={114},
  number={526},
  pages={928--937},
  year={2019},
  publisher={Taylor \& Francis}
}

@article{wen2018sure,
  title={Sure independence screening adjusted for confounding covariates with ultrahigh dimensional data},
  author={Wen, Canhong and Pan, Wenliang and Huang, Mian and Wang, Xueqin},
  journal={Statistica Sinica},
  pages={293--317},
  year={2018},
  publisher={JSTOR}
}

@article{zhang2018robust,
  title={Robust conditional nonparametric independence screening for ultrahigh-dimensional data},
  author={Zhang, Shucong and Pan, Jing and Zhou, Yong},
  journal={Statistics \& Probability Letters},
  volume={143},
  pages={95--101},
  year={2018},
  publisher={Elsevier}
}

@article{liu2014feature,
  title={Feature selection for varying coefficient models with ultrahigh-dimensional covariates},
  author={Liu, Jingyuan and Li, Runze and Wu, Rongling},
  journal={Journal of the American Statistical Association},
  volume={109},
  number={505},
  pages={266--274},
  year={2014},
  publisher={Taylor \& Francis}
}

@article{hong2016data,
  title={A data-driven approach to conditional screening of high-dimensional variables},
  author={Hong, Hyokyoung G and Wang, Lan and He, Xuming},
  journal={Stat},
  volume={5},
  number={1},
  pages={200--212},
  year={2016},
  publisher={Wiley Online Library}
}

@article{liu2018quantile,
  title={Quantile screening for ultra-high-dimensional heterogeneous data conditional on some variables},
  author={Liu, Yi and Chen, Xiaolin},
  journal={Journal of Statistical Computation and Simulation},
  volume={88},
  number={2},
  pages={329--342},
  year={2018},
  publisher={Taylor \& Francis}
}

@article{chen2019simple,
  title={A simple model-free survival conditional feature screening},
  author={Chen, Xiaolin and Zhang, Yahui and Chen, Xiaojing and Liu, Yi},
  journal={Statistics \& Probability Letters},
  volume={146},
  pages={156--160},
  year={2019},
  publisher={Elsevier}
}

@article{zheng2020building,
  title={Building generalized linear models with ultrahigh dimensional features: A sequentially conditional approach},
  author={Zheng, Qi and Hong, Hyokyoung G and Li, Yi},
  journal={Biometrics},
  volume={76},
  number={1},
  pages={47--60},
  year={2020},
  publisher={Wiley Online Library}
}

@article{lu2019conditional,
  title={Conditional distance correlation sure independence screening for ultra-high dimensional survival data},
  author={Lu, Shuiyun and Chen, Xiaolin and Wang, Hong},
  journal={Communications in Statistics-Theory and Methods},
  pages={1--18},
  year={2019},
  publisher={Taylor \& Francis}
}

@article{coles2011proteoglycan,
  title={Proteoglycan-specific molecular switch for RPTP$\sigma$ clustering and neuronal extension},
  author={Coles, Charlotte H and Shen, Yingjie and Tenney, Alan P and Siebold, Christian and Sutton, Geoffrey C and Lu, Weixian and Gallagher, John T and Jones, E Yvonne and Flanagan, John G and Aricescu, A Radu},
  journal={Science},
  volume={332},
  number={6028},
  pages={484--488},
  year={2011},
  publisher={American Association for the Advancement of Science}
}

@article{pulido1995lar,
  title={The LAR/PTP delta/PTP sigma subfamily of transmembrane protein-tyrosine-phosphatases: multiple human LAR, PTP delta, and PTP sigma isoforms are expressed in a tissue-specific manner and associate with the LAR-interacting protein LIP. 1.},
  author={Pulido, Rafael and Serra-Pages, Carles and Tang, May and Streuli, Michel},
  journal={Proceedings of the National Academy of Sciences},
  volume={92},
  number={25},
  pages={11686--11690},
  year={1995},
  publisher={National Acad Sciences}
}

@article{tanaka2015developmental,
  title={Developmental changes in the corpus callosum from infancy to early adulthood: a structural magnetic resonance imaging study},
  author={Tanaka-Arakawa, Megumi M and Matsui, Mie and Tanaka, Chiaki and Uematsu, Akiko and Uda, Satoshi and Miura, Kayoko and Sakai, Tomoko and Noguchi, Kyo},
  journal={PloS one},
  volume={10},
  number={3},
  pages={e0118760},
  year={2015},
  publisher={Public Library of Science San Francisco, CA USA}
}

@book{kennedy2016micro,
  title={Micro-, meso-and macro-connectomics of the brain},
  author={Kennedy, Henry and Van Essen, David C and Christen, Yves},
  year={2016},
  publisher={Springer Nature}
}

@article{bachman2014corpus,
  title={Corpus callosum shape and size changes in early Alzheimer's disease: A longitudinal MRI study using the OASIS brain database},
  author={Bachman, Alvin H and Lee, Sang Han and Sidtis, John J and Ardekani, Babak A},
  journal={Journal of Alzheimer's Disease},
  volume={39},
  number={1},
  pages={71--78},
  year={2014},
  publisher={IOS Press}
}

@article{vermeulen2023morphological,
  title={A morphological study of the shape of the corpus callosum in normal, schizophrenic and bipolar patients},
  author={Vermeulen, Christiaan L and du Toit, Peet J and Venter, Gerda and Human-Baron, Rene},
  journal={Journal of anatomy},
  volume={242},
  number={2},
  pages={153--163},
  year={2023},
  publisher={Wiley Online Library}
}


\end{document}